\documentclass[aps,prl,floats,twocolumn,superscriptaddress,floatfix,showpacs]{revtex4}
\usepackage{epsf,graphicx}
\usepackage{amssymb}
\usepackage{amsmath}
\usepackage{eepic}
\usepackage[dvips]{color}
\newlength{\piclen}
\begin{document}

\title{Quantum critical scaling and superconductivity in heavy electron materials}
\author{Yi-feng Yang}
\email[]{yifeng@iphy.ac.cn}
\affiliation{Beijing National Laboratory for Condensed Matter Physics and \\
Institute of Physics, Chinese Academy of Sciences, Beijing 100190, China}
\affiliation{Collaborative Innovation Center of Quantum Matter, Beijing 100190, China}
\author{David Pines}
\affiliation{Department of Physics, University of California, Davis, California 95616, USA}
\affiliation{Santa Fe Institute, Santa Fe, NM 87501, USA}
\author{N. J. Curro}
\affiliation{Department of Physics, University of California, Davis, California 95616, USA}
\date{\today}

\begin{abstract}
We use the two fluid model to determine the conditions under which the nuclear spin-lattice lattice relaxation rate, $T_1$, of candidate heavy quantum critical superconductors can exhibit scaling behavior and find that it can occur if and only if their "hidden" quantum critical spin fluctuations give rise to a temperature-independent intrinsic heavy electron spin-lattice relaxation rate. The resulting scaling of $T_1$ with the strength of the heavy electron component and the coherence temperature, $T^*$,  provides a simple test for their presence at pressures at which the superconducting transition temperature, $T_c$, is maximum and is proportional to $T^*$. These findings support the previously noted partial scaling of the spin-lattice relaxation rate with $T_c$ in a number of important heavy electron materials and provide additional evidence that in these materials their optimal superconductivity originates in the quantum critical spin fluctuations associated with a nearby phase transition from partially localized to fully itinerant quasiparticles. 
\end{abstract}


\maketitle

\begin{figure}[b]
\centerline{{\includegraphics[width=.5\textwidth]{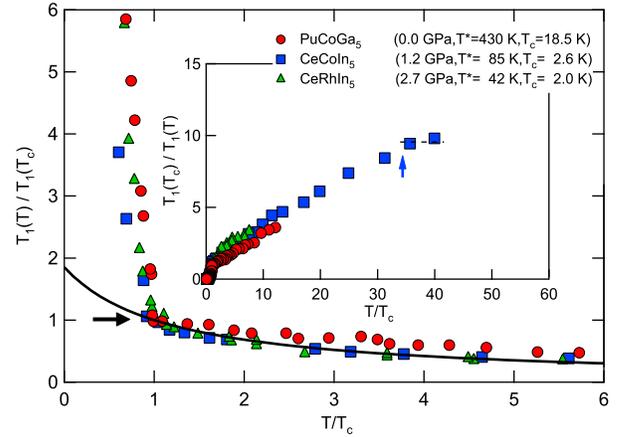}}}
\caption{
{(Color online) The scaling of the spin-lattice relaxation rate with temperature normalized by the superconducting transition temperature for a number of heavy electron superconductors at pressures near that at which $T_c$ is maximum \cite{Curro2005,Kohori2006}. The solid line represents an approximate scaling relation for $T_c<T<T^*/2$, see Eq.~(\ref{Eq:Eq6}). The analysis is extended to higher temperatures in the insert, where the arrows indicate $T^*$. The pressure, $p_L$, at which $T_c$ is maximum is not known for PuCoGa$_5$; it is 1.4 GPa for CeCoIn$_5$ and 2.4 GPa for CeRhIn$_5$, values that differ somewhat  from those for which data is currently available.}
\label{Fig:fig1}}
\end{figure}

\begin{figure}[t]
\centerline{{\includegraphics[width=.5\textwidth]{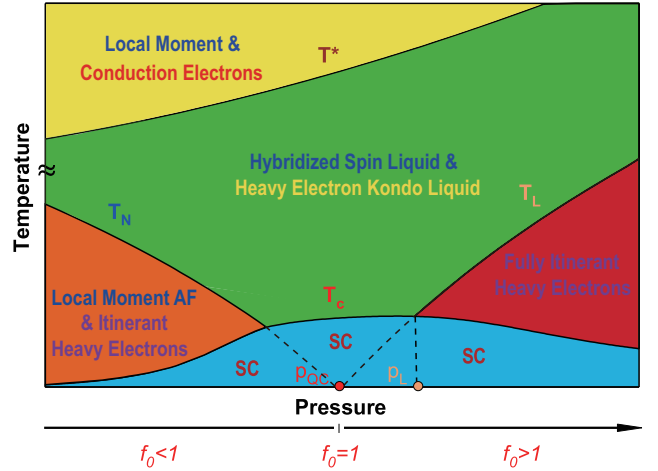}}}
\caption{
{(Color online) A schematic phase diagram for heavy electron superconductors that is based on the application of the two-fluid model to the pressure-induced changes in the behavior of CeCoIn$_5$ and CeRhIn$_5$ \cite{Nicklas2001,Park2008}. It shows the coherence temperature, $T^*$ $(>30 T_c)$, at which itinerant heavy electrons emerge to form a Kondo liquid, and the delocalization line, $T_L$, that marks the boundary between fully itinerant and partially localized heavy electron quasiparticles. Note that $T_c$ is maximum at the pressure, $p_L$, at which $T_c =T_L$, and, as shown in the text, this provides a link between $T_c$ and $T^*$. Importantly, at $p_L$, between $T^*$ and $T_c$, both the hybridized local spin liquid and the itinerant Kondo liquid contribute to the spin lattice relaxation rate.}
\label{Fig:fig2}}
\end{figure}

A tantalizing hint that the spin fluctuations seen in the nuclear spin relaxation rate for a number of unconventional superconductors might be the magnetic glue responsible for their superconductivity appears in a scaling relation between that rate and the optimal superconducting transition temperature, $T_c$, that was first noted by Curro {\sl et al} \cite{Curro2005}. In the present communication we focus on understanding this scaling relation for one important member of this family, the heavy electron materials, for which some experimental results are given in Fig.~\ref{Fig:fig1} \cite{Curro2005,Kohori2006}. As may be seen in Fig.~\ref{Fig:fig2}, finding such a relation appears at first sight highly problematic because the scaling covers a range of temperatures ($T_c<T<T^*$) in the normal state in which both hybridized localized spins and the itinerant heavy electron Kondo liquid contribute to the spin-lattice relaxation rate. However, we find that rigorous Curro $T_c$ scaling can become possible if three conditions are met: (1) the maximum in $T_c$ occurs at the pressure $p_L$, at which the line marking the boundary between partially localized and fully itinerant behavior for heavy electron quasiparticles, $T_L$, intersects with $T_c$, so that $T_c^{\max}=T_L(p_L)$;  (2) at $p_L$ the total spin-lattice relaxation rate scales with the coherence temperature, $T^*(p_L)$; and (3), that at $p_L$, the materials possess identical values of the product $(3T_c/ 2T^*)(T_1^{KL}/T_1^{SL})$ where $T_1^{KL}$ and $T_1^{SL}$ are the intrinsic temperature independent spin-lattice relaxation rates of the Kondo and spin liquids respectively. 

If the first condition is met, $T_c$ will scale with the coherence temperature, $T^*(p_L)$ \cite{Pines2013}. The second scaling relation requires that the intrinsic spin-lattice relaxation rate of the heavy electron Kondo liquid be independent of temperature over much of the relevant temperature region, which will be the case if, and only if, it is caused by magnetic quantum critical spin fluctuations \cite{Yang2009}. We present a simple way to test whether the second condition is met, and find that is satisfied for a number of heavy electron superconductors. The third condition tells us that Curro scaling cannot be universal, since since it involves two material-specific parameters, $3T_c/ 2T^*$ and $T_1^{KL}/T_1^{SL}$, but, as may be seen in Fig.~\ref{Fig:fig1}, these ratios are sufficiently similar for a number of materials that Curro scaling is approximately valid.

The phenomenological two-fluid model, which explains so many other aspects of heavy electron behavior \cite{Yang2009,Nakatsuji2004,Curro2004,Yang2008a,Yang2008b,Yang2012,Shirer2012,Yang2013,Yang2014}, is also key to understanding the remarkable scaling of their dynamic behavior. In it, below $T^*$, the collective hybridization of the Kondo lattice of $f$-electron local moments with the background conduction electrons gives rise to an itinerant heavy electron Kondo liquid (KL) of strength (or volume fraction), $f$, that coexists with the hybridized local moment spin liquid, of strength $1-f$, with \cite{Yang2012}
\begin{equation}
f(T,p)=f_0(p)\left(1-\frac{T}{T^*}\right)^{3/2}.
\label{Eq:Eq1}
\end{equation}
The parameter $f_0(p)$ provides a quantitative measure of hybridization effectiveness; as may be seen in Fig.~\ref{Fig:fig2}, it is unity at the delocalization quantum critical point (QCP) that marks a zero temperature transition from partially localized to fully itinerant behavior, and typically increases with increasing pressure for Ce-based compounds. For $f_0>1$, the model enables one to determine the pressure and temperature dependence of the delocalization line \cite{Yang2012}, $T_L$, shown in Fig.~\ref{Fig:fig2}, that begins at the delocalization QCP and marks, at finite temperatures, the end of the collective hybridization process at $f(T_L,p)=1$. It follows from Eq.~(\ref{Eq:Eq1}) that the delocalization temperature takes the form:
\begin{equation}
T_L(p)=T^*(p)\left[1-f_0(p)^{-2/3}\right].
\label{Eq:Eq2}
\end{equation}

Since partial localization ($f<1$) competes with superconductivity, the two-fluid model predicts that the maximum in the superconducting transition temperature, $T_c^{max}$, will be found at the pressure $p_L$, at which the delocalization line, $T_L$, intersects $T_c$; experiments on CeCoIn$_5$ and CeRhIn$_5$ \cite{Nicklas2001,Park2008} show that this is the case for these materials for which $p_L$=1.4 GPa and 2.4 GPa, respectively. For pressures less than $p_L$, partial quasiparticle localization reduces the number of heavy electrons able to become superconducting and so suppresses $T_c$, while for pressures greater than $p_L$, it is physically appealing to assume that the attractive interaction between heavy electron quasiparticles becomes increasingly less effective \cite{Yang2014b}. 

Note that the quantum critical pressure, $p_{QC}\ne p_L$, as shown in Fig.~\ref{Fig:fig2}. However, for candidate quantum critical superconductors, $f_0(p_L)$ is not far from its value at the quantum critical point, $f_0(p_{QC})=1$, so we can obtain a simple relation between $T_c^{max}$ and $T^*(p_L)$,
\begin{equation}
\frac{T_c^{\max}}{T^*(p_L)}=1-f_0(p_L)^{-2/3}\approx\frac{2}{3}\left[f_0(p_L)-1\right].
\label{Eq:Eq3}
\end{equation}
Since $f_0(p_L)$ is a material sensitive parameter, we see that the ratio of $T_c$ to $T^*$ will in general vary from one material to another and that $T_c/T$ scaling will occur only  for materials with comparable values of $f_0(p_L)$. This is roughly the case for the materials shown in Table~\ref{Table}, in which the value of $f_0$ is estimated from other experiments for CeCoIn$_5$, CeRhIn$_5$ and URu$_2$Si$_2$ and assumed to be near unity at the optimal pressure $p_L$ for PuCoGa$_5$.

\begin{figure}[t]
\centerline{{\includegraphics[width=.5\textwidth]{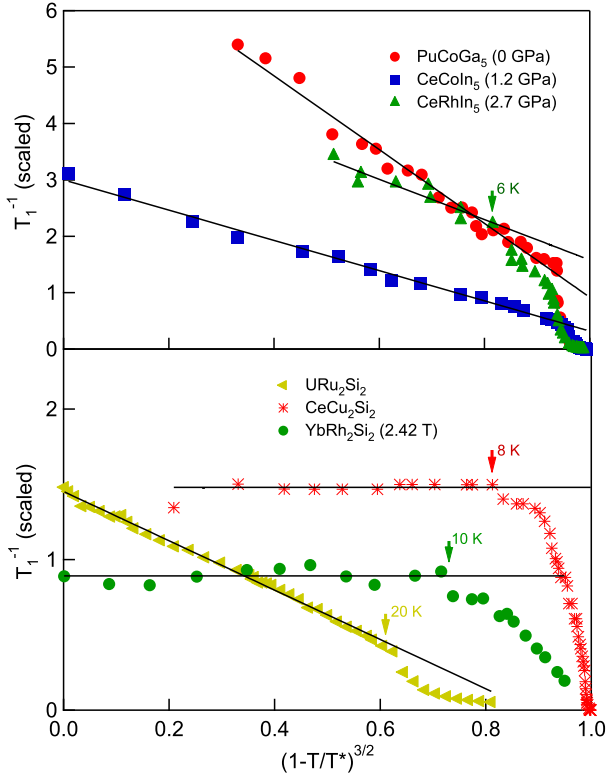}}}
\caption{
{(Color online) Probing quantum critical scaling behavior in a number of heavy electron materials \cite{Curro2005,Kohori2006,Shirer2013,Kitaoka1985,Ishida2002} by plotting the spin-lattice relaxation rate against $(1-T/T^*)^{3/2}$. The arrows indicate the temperature where the scaling fails. Note that temperature decreases from $T^*$ to zero as one goes from right to left on the horizontal axis. While in most of these materials the spin-lattice relaxation rate decreases as the temperature falls below $T^*$ and scales with the strength of the heavy electron component down to some low temperature (indicated by arrows), note that CeCu$_2$Si$_2$ and YbRh$_2$Si$_2$ exhibit markedly different behavior. }
\label{Fig:fig3}}
\end{figure}

We turn next to the two-fluid expression for the spin-lattice relaxation rate \cite{Yang2009},
\begin{equation}
\frac{1}{T_1}=\frac{1-f(T)}{T_1^{SL}}+\frac{f(T)}{T_1^{KL}},
\label{Eq:Eq4}
\end{equation}
where $T_1^{SL}$ and $T_1^{KL}$ are the intrinsic spin-lattice relaxation times of the hybridized local moment spin liquid and the itinerant Kondo liquid. Eq.~(\ref{Eq:Eq4}) tells us that $T/T^*$ scaling will be found if and only if both intrinsic spin-lattice relaxation rates are independent of temperature, with all the measured temperature dependence of $1/T_1$ originating in the strength of the heavy electron component, $f(T)$. For the spin-liquid, experiment suggests that $1/T_1^{SL}$ becomes almost temperature independent as one approaches $T^*$ for materials that are at or near the quantum critical or localization pressure [cf the insert in Fig.~\ref{Fig:fig1}] and it is reasonable to assume that this behavior continues below $T^*$. Importantly, the intrinsic Kondo liquid spin-lattice rate will be nearly independent of temperature over a wide range of temperatures provided the spin fluctuations responsible for its spin-lattice relaxation rate, $1/T_1^{KL}$, exhibit the magnetic quantum critical behavior \cite{Yang2009,Barzykin2009} that has been shown to give rise to  $T_1^{KL}T\sim (T+T_0)$ with $T_0$ being small and going to zero at the QCP.

For any material for which $T^*$ is known, it is straightforward to test the extent to which these conditions are met for the intrinsic spin-lattice relaxation rates: one has only to plot $1/T_1$ vs the strength of the heavy electron component, $(1-T/T^*)^{3/2}$, for temperatures below $T^*$. On rewriting Eq.~(\ref{Eq:Eq4}) as
\begin{equation}
\frac{1}{T_1}=\frac{1}{T_1^{SL}}+\left(\frac{1}{T_1^{KL}}-\frac{1}{T_1^{SL}}\right)f_0(p)\left(1-\frac{T}{T^*}\right)^{3/2},
\end{equation}
we see that to the extent that one finds linear behavior, one can determine directly both $1/T_1^{SL}$ and the product $f_0(p)\left(1/T_1^{KL} -1/T_1^{SL}\right)$ from such a scaling plot. Our results for a number of heavy electron materials are given in Fig.~\ref{Fig:fig3}, where the expected linear scaling is found between $T^*$ and a cut-off temperature, $T_x$, for each material shown there, while the corresponding results for the intrinsic spin lattice relaxation rates are given in Table~\ref{Table}.

We further note that to the extent that the ratio, $T_1^{KL}/T_1^{SL}$, is similar for different superconductors, one can obtain two simple non-universal scaling formulae for the spin-lattice relaxation rate. The first, applicable at temperatures from $T_c$ up to $\sim T^*/2$, concerns the scaling behavior with $T_c$ depicted in Fig.~\ref{Fig:fig1}. Assuming two-fluid scaling behavior persists down to $T_c$, the strength of the heavy electron component is given by $f(T)\approx f_0(p_L)(1-3T/2T^*)\approx 1-3(T-T_c)/2T^*$ at $p_L$, and one easily finds,
\begin{eqnarray}
\frac{T_1(T_c)}{T_1(T)}&=&f(T)+[1-f(T)]\frac{T_1^{KL}}{T_1^{SL}} \nonumber \\
&\approx&1+g^*\left(\frac{T}{T_c}-1\right),
\label{Eq:Eq6}
\end{eqnarray}
where $g^*$ is a non-universal parameter and equal to $(3T_c/2T^*)(T_1^{KL}/T_1^{SL}-1)$, which is calculated and listed in Table~\ref{Table} for each material. A suggested scaling line with $g^*=0.27$ is plotted in Fig.~\ref{Fig:fig1} for comparison with experiment. We see that the overall agreement looks good despite the lack of universality. 

\begin{figure}[t]
\centerline{{\includegraphics[width=.5\textwidth]{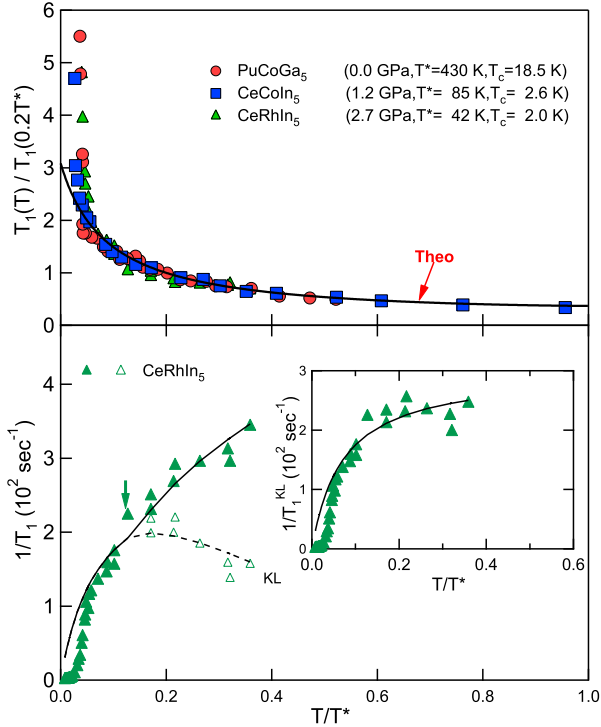}}}
\caption{
{(Color online) Comparison of theory and experiment for the spin-lattice relaxation rate in several heavy electron superconductors. (a) The scaling of $T_1$ as a function of $T/T^*$. The solid line is the proposed scaling formula with $t_0=2.7$. (b) The total $1/T_1$ and the Kondo liquid contribution for CeRhIn$_5$ at 2.7 GPa . The inset shows the intrinsic Kondo liquid term $1/T_1^{KL}$ derived after subtracting the spin liquid contribution. The (solid and dashed) lines are fit to experiment using $T_1^{KL}T\sim (T+T_0)$ with $T_0=3$ K for CeRhIn$_5$.}
\label{Fig:fig4}}
\end{figure}

A second scaling expression emerges if one consider the spin-lattice relaxation rate normalized at a fixed temperature, say $0.2T^*$:
\begin{equation}
\frac{T_1(0.2T^*)}{T_1(T)}\approx t_0+1.4(1-t_0)\left(1-\frac{T}{T^*}\right)^{3/2},
\end{equation}
where $t_0=T_1(0.2T^*)/T_1(T^*)$. As shown in Table~\ref{Table}, $t_0\sim2.7$ for PuCoGa$_5$ and CeCeCoIn$_5$, and is somewhat lower for the other member of the Ce115 family, CeRhIn$_5$. The solid lines in Fig.~\ref{Fig:fig4}(a) shows that $t_0=2.7$ provides a surprisingly good fit to the data for the materials shown there, despite the fact that the hyperfine coupling constants and spin liquid contributions may differ from material to material and not all of these are at the pressure $p_L$ at which $T_c$ is maximum.

We now look into the details of the scaling behavior for individual materials. In Fig.~\ref{Fig:fig3}(a), one sees that the $T^*$ scaling continues down to the superconducting transition in CeCoIn$_5$ ($p=1.2$ GPa) and PuCoGa$_5$. This proves that for CeCoIn$_5$ at $p_L$ the spin fluctuations responsible for the intrinsic Kondo liquid $T_1$ exhibit quantum critical magnetic behavior, and suggests a possibly similar situation in PuCoGa$_5$. We note that recent elastic moduli measurements have observed an anomalous softening of the bulk modulus in PuCoGa$_5$ and suggested that valence fluctuations may play a critical role in its superconductivity \cite{Ramshaw2015}. Combining their results and the NMR measurement \cite{Curro2005} suggests that both valence and spin fluctuations may be involved to give rise to the particularly high $T_c$ in this interesting compound.

For CeRhIn$_5$ at $p=2.7$ GPa, a pressure that is larger than its experimental value of $p_L=2.4$ GPa, we see that scaling behavior is found down to a temperature $T_x\sim T_L$, its delocalization temperature, that is calculated to be 5.7 K using $T^*=42$ K and $f_0$(2.7 GPa)=1.24 estimated from other experiments \cite{Yang2014b}; the loss of scaling below $T_x$ must be attributed to the intrinsic Kondo liquid. As may be seen in Fig.~\ref{Fig:fig4}, a reasonable fit to the experimental data over the entire temperature region may be obtained if we take $T_1^{KL}T\sim (T+T_0)$ with $T_0\sim3$ K providing a measure of the distance away from the quantum critical pressure.

We next examine the scaling behavior, shown in  Fig.~\ref{Fig:fig3}(b), of three other extensively studied heavy electron materials, URu$_2$Si$_2$, CeCu$_2$Si$_2$ and YbRh$_2$Si$_2$. For URu$_2$Si$_2$, we see that as was the case for CeRhIn$_5$ at $p=2.7$ GPa, $T^*$ scaling works down to the delocalization temperature, $T_L\sim20$ K (calculated with earlier estimates of $f_0 =1.6$ \cite{Yang2012}), a temperature that is slightly above the hidden order transition temperature at 17.5 K. This tells us two things: that the hidden order state emerges from a fully formed itinerant heavy electron Kondo liquid; and that its physical origin is likely an attractive quasiparticle interaction induced by their coupling to quantum critical spin fluctuations, an interaction that, in this case, leads to the hidden order state, rather than superconductivity at much lower temperatures. 

The behavior of the spin-lattice relaxation rate of both CeCu$_2$Si$_2$ and YbRh$_2$Si$_2$ at ambient pressure is anomalous. Although a number of experiments have suggested a $T^*\sim70$ K for CeCu$_2$Si$_2$, we see in Fig.~\ref{Fig:fig3}(b) that its spin-lattice relaxation rate $1/T_1$ continues to be temperature-independent down to $\sim8$ K \cite{Kitaoka1985}, below which spin density wave fluctuations are observed in neutron scattering experiments \cite{Stockert2011}. We encounter a similar anomaly in YbRh$_2$Si$_2$ for which $T^*\sim50$ K \cite{Ishida2002,Mo2012}. Assuming the two-fluid model is applicable, this either requires a non-accidental cancellation ($T_1^{SL}\approx T_1^{KL}$) of the independent spin liquid and Kondo liquid contributions to $1/T_1$ over a substantial temperature range ($T\gg T_0$), or a quite different physical picture for spin-lattice relaxation in  these two materials. At this stage, we cannot make explicit predictions on the nature of the QCP in each material.

In summary, we have shown that an analysis based on the two-fluid model leads to a number of non-universal quantum critical scaling relations in the spin-lattice relaxation rate of heavy electron materials at the pressure, $p_L$, at which $T_c$ is maximum and explains the quantum critical spin fluctuation origin of the observed Curro $T/T_c$ scaling. It shows that the total spin-lattice relaxation rate will scale with the strength of the heavy electron component, $f(T/T^*)$, and hence with $T/T^*$ only if quantum critical spin fluctuations determine the intrinsic heavy electron spin-lattice relaxation rate, and yields a simple relation between $T^*$ and $T_c$ at $p_L$. It provides the hitherto missing link between the observation of Curro scaling and its physical origin and enables one to identify materials in which quantum critical spin fluctuations provide the pairing glue. Our results support a strategy for finding higher transition temperature heavy electron superconductors that is based on increasing $T^*(p_L)$ \cite{Pines2013} and raise the interesting question of whether a similar link can be found for other unconventional superconductors for which Curro scaling has been observed.

Y.Y. is supported by the National 973 Project of China (Grant No. 2015CB921303), the National Natural Science Foundation of China (NSFC Grant No. 11174339) and the Strategic Priority Research Program (B) of the Chinese Academy of Sciences (Grant No. XDB07020200). We thank the Aspen Center for Physics (NSF Grant No. 1066293) for their hospitality during the writing of this paper; Y.Y. thanks the Simons Foundation for its support. N.J.C. thanks NNSA de-na0001842-0 for support.

\begin{table*}
\centering
\caption{Results of an analysis of the two-fluid parameters and the spin-lattice relaxation rate in several heavy electron superconductors at indicated pressures. The unit for all the spin-lattice relaxation rates is sec$^{-1}$. $T_x$ is the cut-off temperature below which the scaling breaks down and $t_0=T_1(0.2T^*)/T_1(T^*)$. $f_0(p)$ is determined by Eq.~(\ref{Eq:Eq3}) assuming $p=p_L$ for PuCoGa$_5$ and from previous estimates for all other materials.}
\label{Table}
\begin{tabular}{@{\extracolsep{\fill}}cccccccccccccccc}
\hline\hline
& $p$ (GPa) & $T_c$ (K) & $T^*$ (K) & $T_c/T^*$ & $f_0(p)$ & $1/T_1^{SL}$ (sec$^{-1}$) & $1/T_1^{KL}$ (sec$^{-1}$) & $T_1^{KL}/T_1^{SL}$ & $T_x$ & $t_0$ & $g^*$ & refs. \\ \hline

PuCoGa$_5$ & 0 & 18.5 & 430 & 0.043 & 1.068 & 746  & 134 & 5.6 & $\sim T_c$ & 2.68 & 0.30 & \cite{Curro2005,Pines2013} \\  

CeCoIn$_5$& 1.2 & 2.6 & 85 & 0.031 & 1.023 & 300  & 38 & 7.9 & $\sim T_{c}$ & 2.77 & 0.27 & \cite{Yang2014,Nicklas2001,Kohori2006}\\

CeRhIn$_5$  &  2.7 & 2.0 & 42 & 0.048 & 1.244 & 516  & 228 & 2.3 & $\sim T_L$ & 1.99 & - & \cite{Park2008,Kohori2006} \\ 

URu$_2$Si$_2$ & 0 & 1.5 & 75 & 0.02 & 1.6 & 2.91 & 0.86 & 3.4 & $\sim T_{HO}$ & - & - &\cite{Shirer2013} \\

\hline\hline
\end{tabular}
\end{table*}

\end{document}